# MANAGEMENT OF DATA REPLICATION FOR PC CLUSTER BASED CLOUD STORAGE SYSTEM

Julia Myint[1] and Thinn Thu Naing[2]

[1]University of Computer Studies, Yangon, Myanmar
juliamyint@gmail.com
[2]University of Computer Studies, Yangon, Myanmar
ucsy21@most.gov.mm

## ABSTRACT

*Storage systems are essential building blocks for cloud computing infrastructures. Although high performance storage servers are the ultimate solution for cloud storage, the implementation of inexpensive storage system remains an open issue. To address this problem, the efficient cloud storage system is implemented with inexpensive and commodity computer nodes that are organized into PC cluster based datacenter. Hadoop Distributed File System (HDFS) is an open source cloud based storage platform and designed to be deployed in low-cost hardware. PC Cluster based Cloud Storage System is implemented with HDFS by enhancing replication management scheme. Data objects are distributed and replicated in a cluster of commodity nodes located in the cloud. This system provides optimum replica number as well as weighting and balancing among the storage server nodes. The experimental results show that storage can be balanced depending on the available disk space, expected availability and failure probability of each node in PC cluster.*

## 1. INTRODUCTION

Cloud computing has become a significant technology trend, and many experts expect that cloud computing will reshape information technology (IT) processes and the IT marketplace. With the cloud computing technology, users use a variety of devices, including PCs, laptops, smart phones, and PDAs to access programs, storage, and application-development platforms over the Internet, via services offered by cloud computing providers. Advantages of the cloud computing technology include cost savings, high availability, and easy scalability. [15]

Companies such as Google, Amazon and Microsoft have been building massive data centres over the past few years. Spanning geographic and administrative domains, these data centres tend to be built out of commodity desktops with the total number of computers managed by these companies being in the order of millions. Additionally, the use of virtualization allows a physical node to be presented as a set of virtual nodes resulting in a seemingly inexhaustible set of computational resources. By leveraging economies of scale, these data centres can provision CPU, networking, and storage at substantially reduced prices which in turn underpins the move by many institutions to host their services in the cloud. [15]

Cloud storage (or data storage as a service) is the abstraction of storage behind an interface where the storage can be administered on demand. Further, the interface abstracts the location of the storage such that it is irrelevant whether the storage is local or remote (or hybrid). Cloud storage infrastructures introduce new architectures that support varying levels of service over a potentially large set of users and geographically distributed storage capacity.

This time, however, the economic situation and the advent of new technologies have sparked strong interest in the cloud storage provider model. Cloud storage providers deliver economics of scale by using the same storage capacity to meet the needs of storage user, passing the cost savings to their storage.

Many cloud service providers use high performance storage server as datacenter which is very expensive and reliable. However, storing information and managing its storage in a limited budget is a critical issue for a small business as well as for large enterprises. Vendors come up with different solutions day by day but these solutions are very expensive and hard to maintain.

In this paper, we propose a solution that addresses the above mentioned problems. Nowadays, a standard desktop PC has enormous computing and storage capacity. Usually, a standard PC contains more than 40 GB hard disk drive (HDD), 256 MB RAM and 2 GHz or higher processor. If the available storage capacity of these PCs is combined together, then a PC cluster containing 20 PCs with above mentioned specifications can provide 20x40 =800 GB of storage capacity. Example of storage system mentioned above is shown in figure 1.

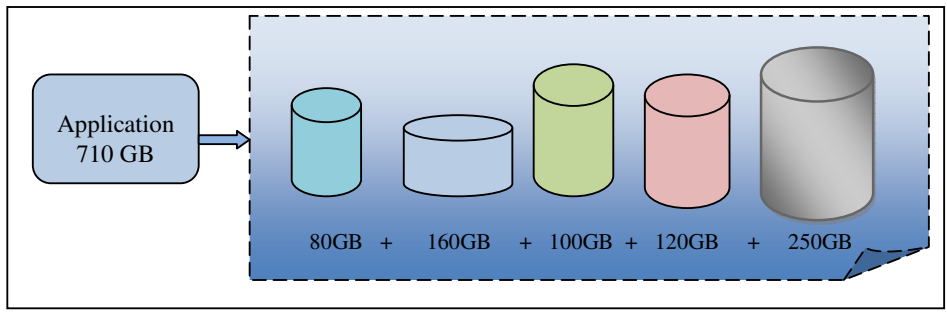

Figure 1.  Example of large scale data storage on heterogeneous nodes

In cluster technology, many computers or storage nodes are connected together to provide high capacity storage server. But storage nodes are connected in a cluster may result in obvious load balancing problems and scalability problems.

This paper proposes the design of cloud storage system which utilizes a PC cluster consisting of computer machines (PCs) operating in our university. This solution is very cost effective because any organization or university can utilize this system over their existing resources (desktop machines) without purchasing any extra hardware or software components.

The rest of this paper is organized as follows. In the next section, we discuss the related papers with our paper. In section 3, we demonstrate infrastructure of proposed cloud storage on PC cluster. And then we describe the replication management and replica placement algorithm in section 4. Then data loss probability is analyzed in section 5. In section 6, we discuss experimental evaluation. Finally, we conclude our paper.

## 2. RELATED WORK

There are large amount of researches in the design of cloud storage. Many of these researches are file system based storage system such as GFS[13] and HDFS[6]. These architectures are master-slave routing paradigm. In those storage systems, replication management is performed by using default replica number. Moreover, the load balancing is achieved by data migration in these systems. It can cost more bandwidth utilization to the whole system.

Hoang Tam Vo and et al. [8] proposed ecStore, a highly elastic distributed storage system with range-query and transaction support. The design of ecStore is based on distributed index structure called BATON (Balanced Tree Overlay Network) [9]. Despite the effect of single master node failure, it is organized into peer-to-peer virtual network structure on storage nodes. The ecStore also applied data migration technique to balance the load of storage nodes and it needs to restructure the BATON whenever load unbalanced. Other cloud storage systems that use data migration for load balancing are Dynamo [7], Pnuts[5] and Cassandra[3].

Data placement based on replication management has been active research issue in storage system. Qingsong Wei [12] proposed a cost-effective replication management scheme for cloud storage cluster. In that paper, blocking probability is used as a criterion to place replicas among data nodes to reduce access skew, so as to improve load balance and parallelism. In [14], cloud storage is designed based on hybrid of replication and data partitioning. Two level DHT approach is proposed for widely distributed cloud storage.

The above cloud storage systems apply different strategies for effective storage, but they do not consider available storage loads when data is placed to the storage server. In this paper, therefore, an efficient replication management scheme is proposed with the support of availability and load balancing over PC cluster.

## 3. PC Cluster based Cloud storage architecture

Cloud storage infrastructure is deployed in PC cluster. Since we apply PC nodes as cloud storage server instead of high performance storage server, there are many challenges such as

- Faster access to the distributed files
- Fault-tolerance
- Load balancing
- Availability
- Automated data partitioning and replication

To address these problems, we propose cloud storage infrastructure as shown in figure2.

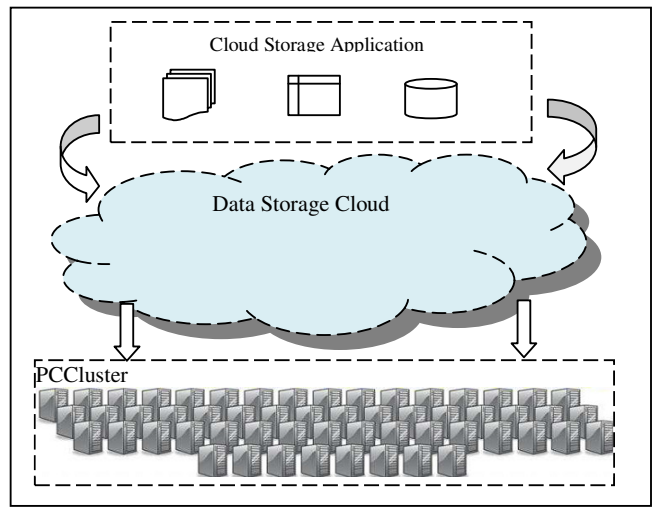

Figure 2. Overview of the cloud storage infrastructure

### 3.1. Cloud Application

In cloud technology, users browse applications through browsing interfaces. The applications may involve data storage and file retrieval applications. Especially, it will be deployed and integrated with other cloud based open source application such as AppScale[11] and Hbase [2]. Then the data objects storage requests will be sent to web servers and forwarded them to cloud storage clients through a request controller. Then, clients interact with storage servers called PC cluster to fetch requested files.

## 3.2. Cloud Storage Client

Cloud storage client is an interface between user interface and storage servers. User applications access the storage servers using the storage client which is a code library that exports the cloud storage client. The process of accessing the storage servers is shown in figure 3.When an application reads a file, the storage client first asks the Name node for the list of Data nodes that host replicas of the blocks of the file. It then contacts a Data node directly and requests the transfer of the desired block. When a client writes, it first asks the Name node to choose Data nodes to host replicas.

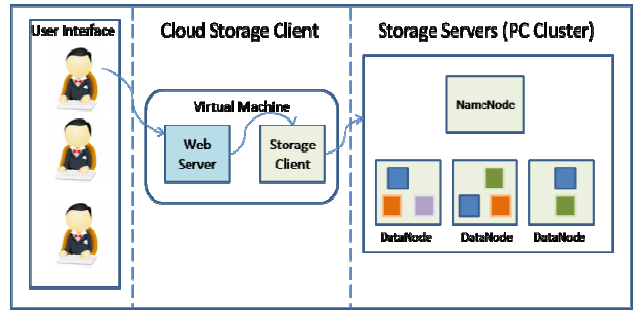

Figure 3.  Access process in cloud storage infrastructure

## 3.3. PC Cluster

Clusters have networks with low latency and high bandwidth, and large numbers of trusted, centrally-managed hosts. In addition, the opportunity cost of wasted resources is potentially very high, so cluster systems must make good use of the combined storage capacity and I/O bandwidth available from the array of component machines.

The designed storage system is based on PC cluster for cloud. PC cluster is used for data storage. Data on cloud computing is stored in PC cluster. The storage system consists of a single Name node and a number of Data nodes, usually one per node in the PC cluster, which manages storage attached to the nodes that they run on. The PC cluster-based cloud storage system is shown in figure 4.

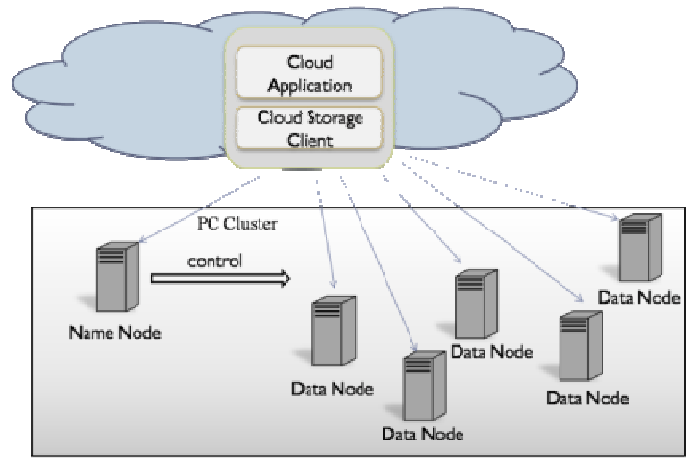

Figure 4.  PC cluster-based cloud storage system

In this system, the Name node manages the whole PC cluster. It also executes file system namespace operations like opening, closing, and renaming files and directories. It also determines the mapping of blocks to Data nodes. The Data nodes are responsible for serving read and write requests from the clients. The Data nodes also perform block creation, deletion and replication upon instruction from the Name node.

## 4. DATA PLACEMENT MANAGEMENT

Data placement is critical to storage system for data availability and fault tolerance. A good data placement policy should improve data reliability, availability and network bandwidth utilization. Therefore, we focus replication strategy to achieve these goals.

### 4.1. System Model

The cloud storage system is implemented with PC cluster system. It is composed of N independent heterogeneous nodes storing of M different blocks $b_1, b_2, \ldots, b_M$. We define failure probability of data node $S_i$ as $f_i$, the optimum replica number of a data block as $R_{opt}$. The availability of the system, $\alpha$ is defined as the fraction of time that the system is available for serving user requests. The system can be in one of the two states: the normal state or the idle state.

In normal state, at least one data node is available for serving user requests. However, idle state, $P_0$, is the state in which all data nodes containing the data block $b_j$ have failed.

### 4.2. Optimum Replica Number

The goal of replication is to increase reliability and availability by keeping the data accessible even when failures occur in the system. It is clear that the reliability of a system will generally increase as the number of replicas or the degree of replication increases since more replicas will be able to mask more failures. However, it is a key issue how the degree of replication will affect system availability [14-16].

When replica number reaches a certain point, the file availability is equal to 1, and adding more replicas will not improve the file availability any more. The lower the node failure ratio, the less replica number it requires for file availability to reach 1. Therefore, we can only maintain minimum replicas to ensure a given availability requirement [16].

With replica number increasing, the management cost including storage and network bandwidth will significantly increase. In a cloud storage system, the network bandwidth resource is very limited and crucial to overall performance. Too much replicas may not significantly improve availability, but bring unnecessary spending instead.

With attention to this, we developed a model to express availability as function of replica number which is adapted from primary site approach described in [16]. This model is used to determine how much minimal replica should be maintained to satisfy availability requirement. Since the availability of the system is opposite of the idle state of the system, we get

$$\alpha = (1 - P_0) \tag{1}$$

The idle state of the system $P_0$ means that all replica nodes have failed with failure probability fi. That is

$$P_0 = f_1 \times f_2 \times f_{13} \times \ldots \ldots \times f_k = \Pi_{i=1}^{R_n} f_i \tag{2}$$

Therefore, if the availability probability $\alpha$ and failure probability $f_i$ is known, we get the optimum replica number $R_{opt}$ by using the following equation such as:

$$\alpha \leq \min(1 - \Pi_{i=1}^{R0} f_i, \ 1 - \Pi_{i=1}^{R1} f_i, \ldots, 1 - \Pi_{i=1}^{Rn} f_i) \tag{3}$$

According to the equation (3), Name node calculates optimum replica number $R_{opt}$ to satisfy expected availability with average data node failure probability $f_i$ .In case of data node failure and current replica number of a block is less than $R_{opt}$ , additional replicas will be created into the cluster.

### 4.3. Node Weighting

In the proposed system, there is an assumption that each data node has its own specifications such as memory, CPU, disk space, failure probability. Therefore, the proposed strategy takes these specifications into account to choose the Data nodes for storage. Each node has its own weight which is the function of the total storage capacity and node availability shown in equation 4 and 5.

$$DN_{weight} = DN_{DiskSpace} \times DN_{avail} \qquad (4)$$

$$DN_{avail} = 1 - DN_f \qquad (5)$$

In equation 4 and 5, $DN_{weight}$ means the weight of Data node and $DN_{DiskSpace}$ and $DN_{avail}$ imply the storage capacity and availability respectively. In equation 5, $DN_f$ is failure probability of that Data node. For example, if the capacity of drive in PC is 80 gigabytes and failure probability is 0.2, the weight of that PC node is 64.

### 4.4. Replica Placement Algorithm

After determining how many replicas the system should maintain to satisfy availability requirement, we will consider how to place these replicas efficiently to maximize performance and load balancing. In this paper, efficient data placement algorithm is used to place replicas among data nodes to improve load balance. The replica placement algorithm is shown in figure 4.

---

**ReplicaPlacement Algorithm**

**Input** : failure probability $P_f$, availability **α**, data block **b**
**Output** : List of DataNodes **DN**
ReplicaCount = Calculate using equation (3)
i ⟵ 0
SelectedNodes = null
While (i < ReplicaCount)
   DN[i] = choose largest node in PC cluster
   Calculate weight of each node using equation (4) and (5)
   Remove selected node **DN[i]**
   i ⟵ i+1
End While
Return **DN**

---

## 5. ANALYSIS OF DATA LOSS PROBABILITY

The probability of data loss is considered in a cluster to analyze the proposed system. In this paper, the probability of data loss is considered by varying the parameters shown in table I.

Table 1. Parameters used in analysis.

| Parameters | Description |
|---|---|
| p | Probability of failure of a single machine |
| n | Number of machines in the cluster |

| r | Number of replicas for each block |
|---|---|
| b | Total number of data blocks in the cluster |
| α | Expected availability |

In the proposed system, since the system is designed for a storage system using inexpensive computer machines, these machines may fail down. We assume machine failures are independent with failure probability. In our analysis, data loss probability of a PC cluster is modelled as equation 6.

$$P_{loss(n,b,r)} = 1 - \sum_{f=0}^{n} \binom{n}{f} * p^f * (1-p)^{(n-f)} * (1 - \binom{f}{r} / \binom{n}{r})^b \qquad (6)$$

To analyze the data loss probability of proposed system, equation 6 is used with the variation of four parameters n, p, α, r. In the proposed system, replica number r is computed by using equation 3. For simplicity, we assume that each machine has 80 GB of disk space and uses block size of 64 MB and thus each machine has $1280/r_{avg}$ blocks and we also assume that $r_{avg}$ is 3. The results of analysis are shown in table 2.

Table 2. Probability of data loss for some sample values **n, α, p and r**.

| n | α | p | r | Probability of data loss |
|---|---|---|---|---|
| 10 | 0.99 | 0.01 | 2 | 0.0043 |
| | | 0.1 | 3 | 0.0686 |
| | | 0.2 | 3 | 0.3165 |
| 30 | 0.99 | 0.01 | 2 | 0.0286 |
| | | 0.1 | 3 | 0.2179 |
| | | 0.2 | 3 | 0.7120 |
| 60 | 0.99 | 0.01 | 2 | 0.0456 |
| | | 0.1 | 3 | 0.2752 |
| | | 0.2 | 3 | 0.8377 |

According to table 2, probability of data loss mainly depends on failure probability and replica number. Moreover, the probability increases with cluster size. Therefore, proposed data placement algorithm with optimum replica number is more efficient on PC cluster based storage system.

## 6. SYSTEM EVALUATION

The proposed cloud storage system is implemented with Hadoop storage cluster (HDFS). HDFS is an open source storage platform and designed to be deployed in low-cost hardware. Therefore, it is suitable to implement and evaluate the proposed system. In HDFS, the replication factor is configurable per file and can be changed later. However, HDFS does not provide policy to determine the replication factor. In proposed replication management, optimum replica number is determined before data is stored to storage servers and the optimum replica number can recover not only failure probability but also expected availability.

### 6.1. Cluster Setup

The tested cluster consists of five heterogeneous nodes, whose parameters are summarized in Table III. All nodes are connected with 1 Gbps Ethernet network. In each node, Ubuntu server 10.10 is installed. Java version is 1.6.2 and Hadoop version is 0.20.2. The size of HDFS block is 64 MB.

Table 3.Five nodes in PC cluster.

| Node | CPU Model | CPU (Hz) | Memory | Disk Space |
|------|-----------|----------|--------|------------|
| NameNode | Pentium ® Dual-Core | 2.00 GHz | 2GB | 250 GB |
| DataNode1 | Pentium P4 | 1.5 GHz | 512 MB | 80GB |
| DataNode2 | Pentium P4 | 1.5 GHz | 512 MB | 80GB |
| DataNode3 | Pentium P4 | 1.5 GHz | 512 MB | 80GB |
| DataNode4 | Pentium P4 | 1.5 GHz | 512 MB | 80GB |

### 6.2. Data Distribution over the Cluster

The original HDFS is evaluated with the proposed storage cluster. During the evaluation, 10 files are stored in the cluster. Each file has 1000 MB size and the total storage is about 9.765625 GB. The figure 6 shows the distribution of storage in the cluster when the replication factor is set to 1, 2 and 3.

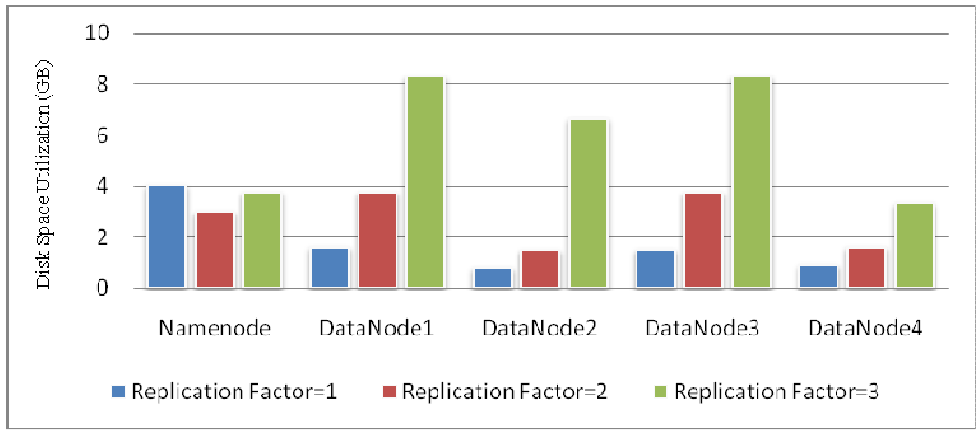

Figure 6: Data distribution of HDFS cluster

### 6.3. Storage Utilization over the Cluster

Storage utilization is defined as the percentage of storage over total storage capacity of data node. The percentage of disk space utilization is computed as equation 7.

$$DSU(\%) = UDS\ (GB) \times \left(\frac{100}{DN_{DiskSpace}}\right) \qquad (7)$$

In equation 7, DSU (%) means the percentage of disk space utilization; UDS (GB) is defined as disk space utilization in gigabytes and $DN_{DiskSpace}$ as initial disk size of Data node.

To evaluate storage utilization, the numbers of replication factor is set to 1, 2 and 3 in original HDFS cluster. For the proposed algorithm, the process is simulated in the experiments and we set the expected availability as 0.99 and the probability of five nodes in the cluster as 0.01, 0.01, 0.02, 0.03 and 0.04 respectively. Moreover, 10 files with total size of about 9 GB are stored to the cluster. The results of experiment are shown in figure 7.

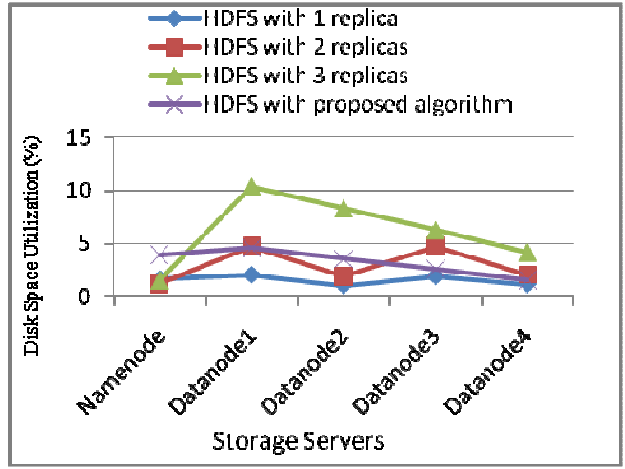

Figure 7: Storage utilization of HDFS cluster

According to the results of figure 7, in the proposed algorithm, disk space utilization is more stabilized depending on the disk size, expected availability and failure probability of each node in the cluster. Moreover, the proposed system can save storage costs because of optimum replica number.

## 6.4. Load Balance

In order to achieve the load balance of the PC cluster, we apply node weighting over PC cluster. The proposed replica placement using node weighting makes the storage loads of the nodes to be balanced when the data block is added. To evaluate load balancing of the cluster, we compute a value which is a fraction in the range of (0, 1). A cluster is more balanced if for each Data node, the value is close to 1.It means that the utilization of the node less is less different from the utilization of the whole cluster. To get the value of load balance in PC cluster, we compute as equation 8.

$$LB_{DN} = 1 - \left| \left( \frac{UDS_{UN}}{total_{DN}} \right) - \left( \frac{DS_{cluster}}{total_{cluster}} \right) \right| \qquad (8)$$

In equation 8, $LB_{DN}$ means load balancing value of Data node DN, $UDS_{DN}$ is defined as used disk space of Data node and $total_{DN}$ is total capacity of Data node. For cluster, $DS_{cluster}$ is defined as used disk space of the whole cluster and $total_{cluster}$ means total capacity of the whole cluster. To compare original HDFS with proposed system, we set up availability as 0.99 and failure probability as 0.01 for each Data node. Figure 8 shows the comparisons of different load balance on PC cluster. With our experiment set up, proposed system and replication factor 1 for HDFS result in a higher load balance.

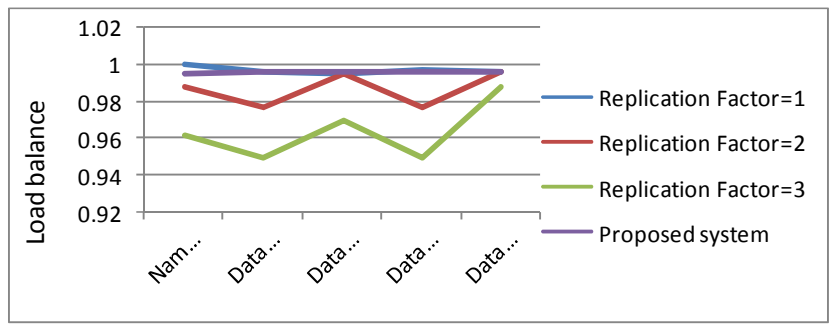

Figure 8: Comparison of load Balance on PC cluster

## 6.5. Effect of Varying Failure Probability

In section 5, we demonstrated that the proposed replication management depends on the failure probability of nodes. Therefore, to evaluate the effects of failure probability, we set up four Data nodes with disk space 80 GB each. To evaluate the proposed system, failure probability of Data node1 is changed from 0.01 to 0.1 while other Data nodes are set to be stable at 0.05. The results of varying failure probability are shown in figure 9. The results of figure 9 prove that the proposed system can adjust the storage distribution depending on the failure probability.

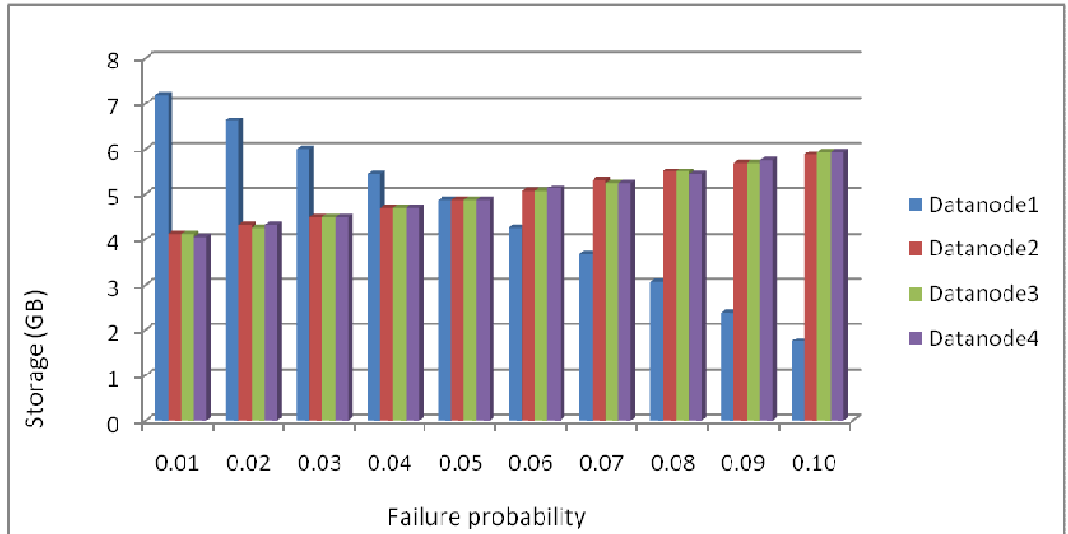

Figure 9: Storage distribution for varying failure probability

## 7. CONCLUSIONS

In this paper, we proposed a cloud storage design on PC cluster. As the system is designed for a storage system using inexpensive computer machines, these machines may fail down and load unbalance which can be slow down and longer data retrieval time. Consequently, we propose replication management scheme for choosing optimum replica number and load balancing. As can be seen from experimental results, the proposed replication management scheme is able to balance the load of PC cluster when data is stored. In the future, we intend to enhance the performance of the system and believe that the proposed system will be cost effective and efficient utilization of computer machines.

## ACKNOWLEDGEMENTS

The authors would like to thank everyone, just everyone!

2008.

**Authors**

**Julia Myint.** She received the Bachelor of Computer Science degree and Master of Computer Science degree from University of Computer Studies, Yangon, Myanmar in 2004 and 2008 respectively. Currently she is a Ph.D. student in University of Computer Studies, Yangon, Myanmar.  Her research interests include Distributed Computing and Cloud Computing.

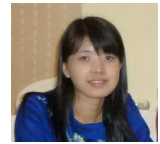

**Thinn Thu Naing.** She obtained her Ph.D degree in Computer Science from University of Computer Studies, Yangon in 2004, Bachelor of Computer Science degree and Master of Computer Science degree in 1994 and 1997respectively from University of Computer Studies, Yangon. Currently, she is a Professor in Computer Science at University of Computer Studies, Yangon, Myanmar.  Her specialization includes cluster computing, grid computing, cloud computing and distributed computing.

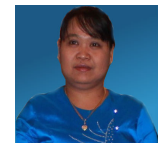